\documentclass[times]{akari2017} 

\shorttitle{Galaxy clustering in the AKARI NEP-Deep}
\shortauthors{A. Pollo, A. Solarz et al.}

\begin{document}

\title{AKARI NEP-Deep: galaxy clustering through the AKARI IRC filters} 



\correspondingauthor{Agnieszka Pollo}
\email{agnieszka.pollo@ncbj.gov.pl}

\author{Agnieszka Pollo} 
\affiliation{National Center for Nuclear Research} 
\affiliation{Astronomical Observatory of the Jagiellonian University} 
\author{Aleksandra Solarz} 
\affiliation{National Center for Nuclear Research} 
\author{AKARI NEP-Deep team} 



\begin{abstract}
We present a preliminary analysis of clustering of galaxies luminous in the near- and mid-infrared as seen by seven various filters of the AKARI IRC instrument from 2$\mu$m to 24$\mu$m in the the AKARI NEP-Deep field. We compare populations of galaxies detected in different filters and their clustering properties. We conclude that different AKARI filters allow to trace different populations composed mainly of star-forming galaxies located in different environments. In particular, the mid-infrared filters at redshift $z \sim 0.8$ and higher trace a population of strongly evolving galaxies located in massive haloes which might have ended as elliptical galaxies today.
\end{abstract}



\setcounter{page}{1}



\section{Introduction} 

One of the main questions of cosmology is how an almost uniform primordial matter density  field was transformed into today's complex web of different types of galaxies. In a standard hierarchical scenario, it was gravitational instability that caused primordial anisotropies in the distribution of matter - mainly cold dark matter - to grow and cluster, and to form haloes of dark matter which detached from the Hubble flow. However, formation of galaxies in these haloes is driven by a much more complex physics of baryonic matter. As a result, galaxies are only {\it biased} tracers of the underlying dark matter density field, with bias evolving with time differently for different galaxy populations. 

\begin{figure}[!ht]
\begin{center}
\resizebox{0.86\hsize}{!}{
        \includegraphics*{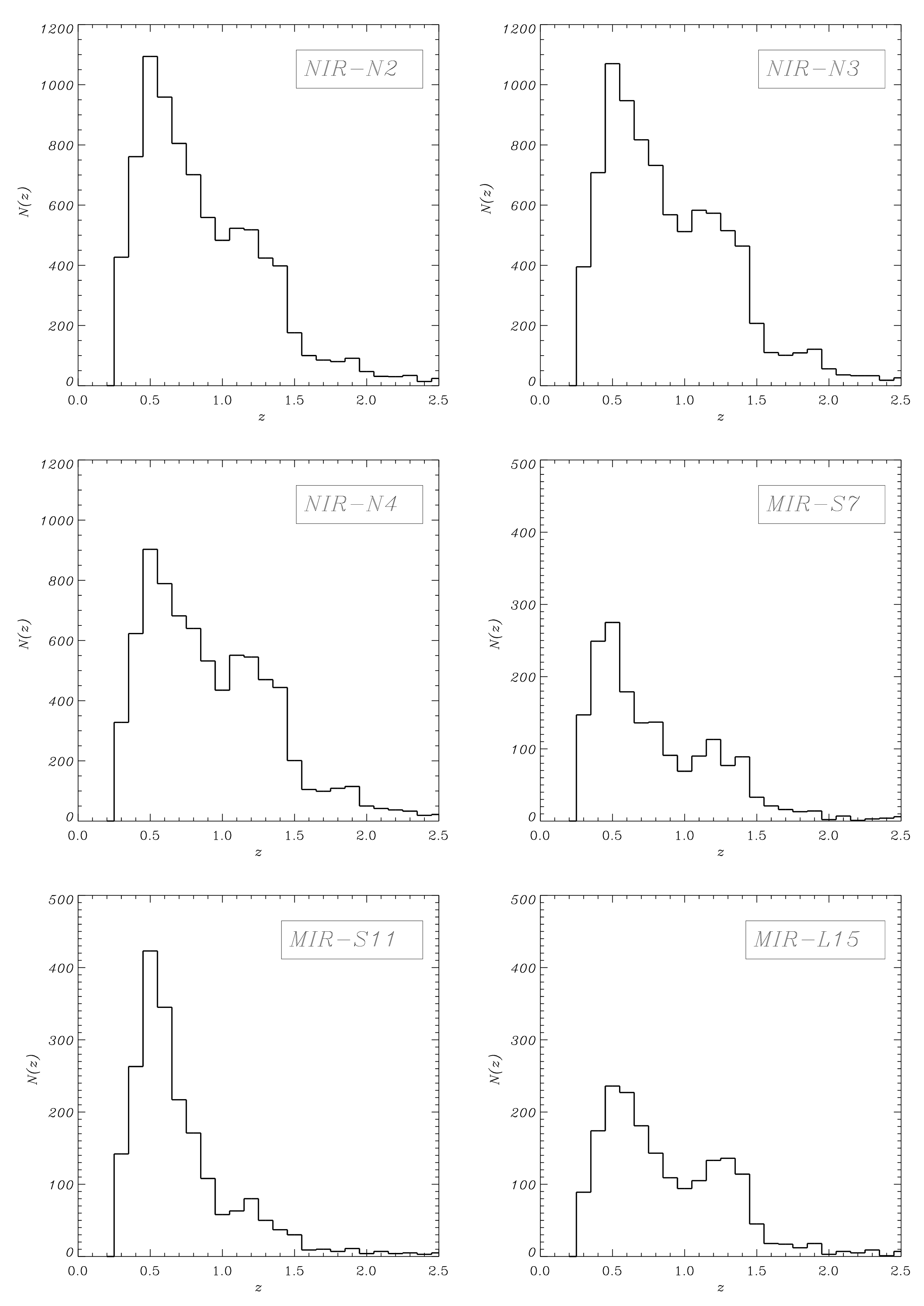}
        }
        \caption{Redshift histograms N(z) of galaxies identified in six IRC filters from 2 $\mu$m to 15 $\mu$m in the AKARI NEP-Deep field .
        }\label{fig:hist}
\end{center}
\end{figure}

The situation is getting even more complex when we realize that observations in optical range do not provide the complete census of galaxies. In particular, dusty star forming galaxies which are very bright in the infrared (IR), can be very dim in the visible light. Of particular interest are near- and mid-infrared (NIR and MIR) observations, as they trace the dust emission resulting from the heating by high-energy photons from young stars. The MIR selection is affected by prominent spectral features, including strong emission from the polycyclic aromatic hydrocarbons (PAH) at rest-frame wavelengths of 3.3, 6.2, 7.7, 8.6, 11.3, 12.7, 16.3, and 17$\mu$m, which can increase the number of sources within a flux-limited sample, or silicate absorption feature at 9.7$\mu$m, which can cause a deficit in detection of sources. Taking all these complex selection effects into account, we need to find out how different populations of star-forming galaxies are placed in the large scale structure in different cosmic epochs in order to understand both their origin and role in the whole galaxy budget. 

Until now, analyses of clustering of MIR-selected galaxies in literature were not very numerous. At shorter wavelengths (3.6$\mu$m, 4.5$\mu$m) \citet{delatorre2007} and \citet{waddington2007} measured clustering of galaxies observed by Spitzer. Measurements on the samples selected at the longer wavelengths were also mainly based on the 24 $\mu$m Spitzer selection  \citep{mgi2008,starikova2012,dolley2014}. \citet{solarz2015} presented the first clustering measurement for the 24 $\mu$m selected sample based on the AKARI North Ecliptic Pole (NEP-Deep) data, finding there three different populations of star-forming galaxies: a sample composed mostly of local star forming galaxies at redshift $z \sim 0.25$, and samples at $z \sim  0.9$ and (tentatively) at $z \sim 2$, dominated by Luminous and Ultra-Luminous Red Galaxies (LIRGs and ULIRGs). Clustering of these samples indicates that while star-forming galaxies at low redshift are mostly field galaxies, the 24$\mu$m selected higher redshift samples contain rather galaxies located in dense environments, likely ancestors of today's elliptical galaxies. We also find that the clustering length and, consequently, dark halo mass is not necessarily correlated with the infrared luminosity of galaxies, in contrast to what is usually observed for optical selection.

\section{Data}

The AKARI NEP-Deep survey covers an area of 0.4 sq. deg around the North Ecliptic Pole. The data were obtained by the AKARI Infrared Camera \citep[IRC][]{onaka2007} through nine NIR and MIR filters, centred at 2$\mu$m (N2), 3$\mu$m (N3), 4$\mu$m (N4), 7$\mu$m (S7), 9$\mu$m (S9W), 11$\mu$m (S11), 15$\mu$m (L15), 18$\mu$m (L18W), and 24$\mu$m (L24).

This study is based on catalogs of galaxies compiled by \citet{murata2013}, completed with photomeric redshifts computed by \citet{oi2014}. We omitted samples selected in broad filters S9W and L18W, as they partially overlap narrow filters, and may be more difficult to interpret in terms of spectral features they follow. The selection gave 13,379 galaxies selected at 2$\mu$m, 14,661 galaxies at 3$\mu$m, 13,113 at 4$\mu$m, 4,479 at 7$\mu$m, 4,650 at 11 $\mu$m, and 2,917 at $\mu$m (of course these samples partially overlap). Additionally, for 24$\mu$m selection we use the results obtained by \citet{solarz2015}. 

\begin{figure}[!ht]{
        \plottwo{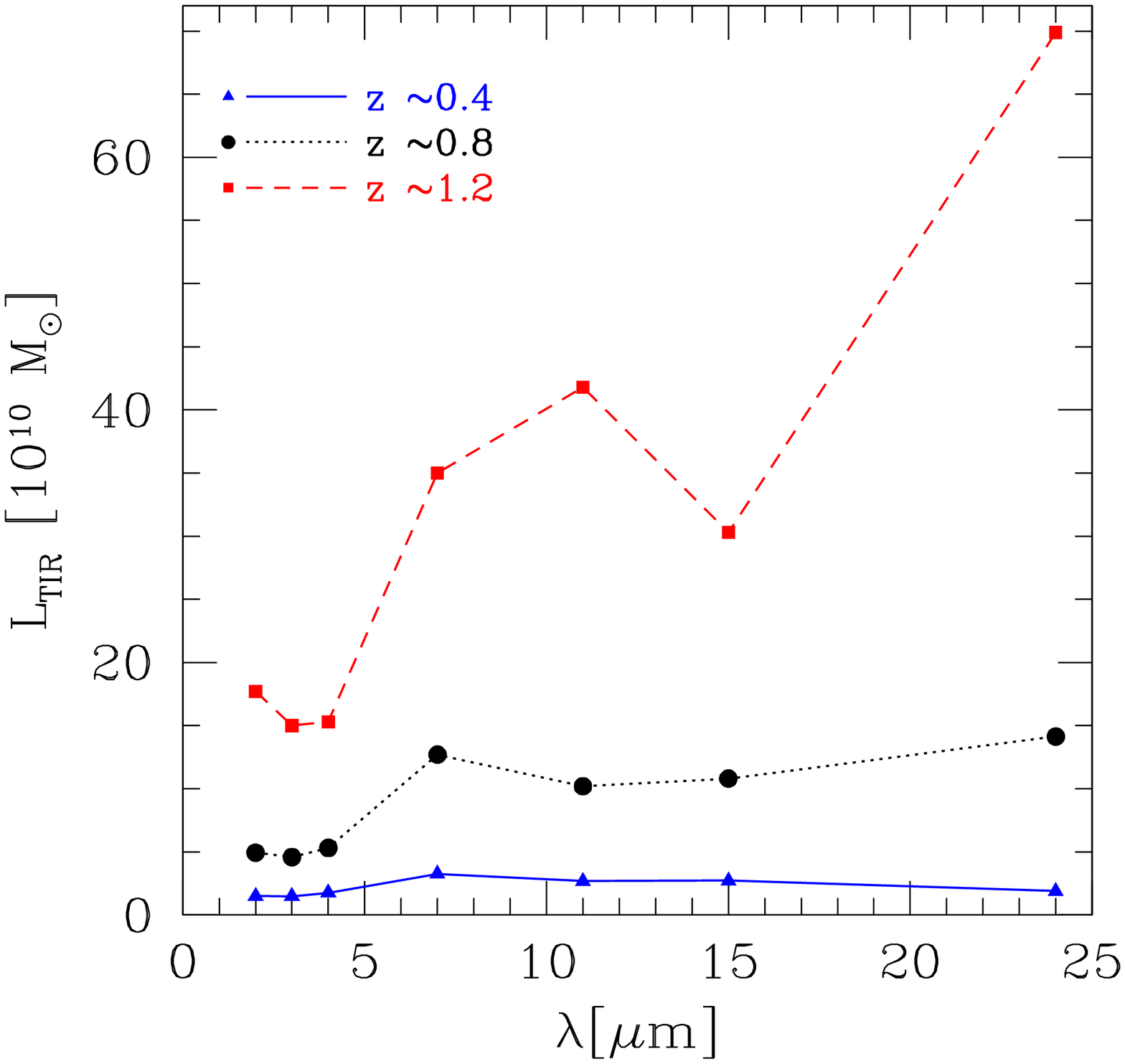}{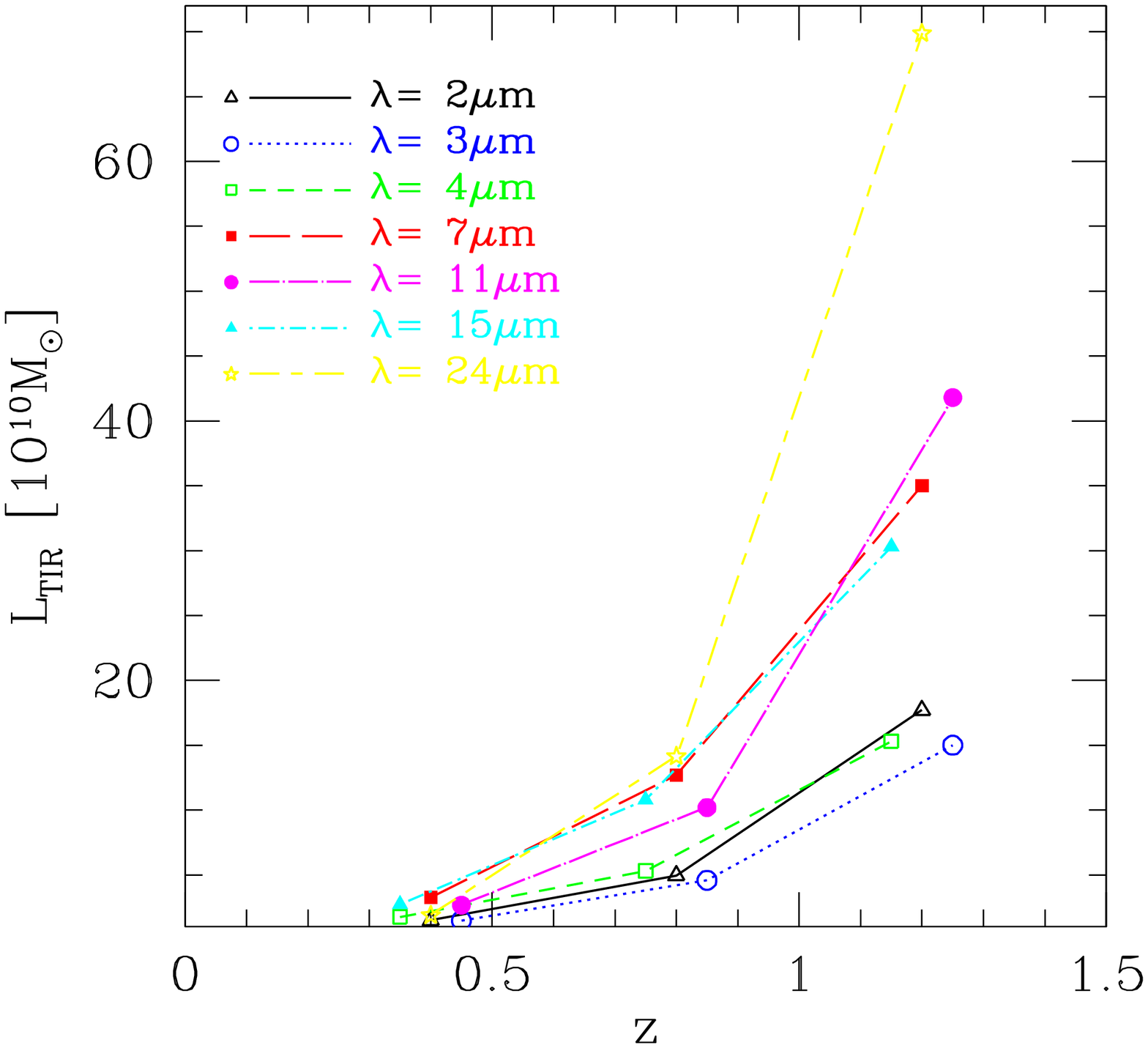}
        }
        \caption{Total median infrared luminosity of samples of AKARI-NEP galaxies in different redshift ranges as a function effective wavelengths of the IRC filter (left panel) and as a function of redshift for samples probed at different redshifts.
        }\label{fig:lum}
\end{figure}

It should be noted, however, that the catalogs mentioned above at 2 to 15$\mu$m contain only objects with optical identifications. This makes them incomplete to a different degree. A comparison with purely IR-selected samples created with the aid of a Support Vector Machine (SVM) based algorithm by \citet{solarz2012} indicates that the most incomplete are the NIR-selected samples (up to 34\%), while the MIR samples, albeit smaller, are usually more complete. This incompleteness may affect to some extend the results of clustering measurements. The 24$\mu$m selected sample used by \citet{solarz2015} which we use for comparison was created with the aid of SVM and we expect it to be complete, but containing sources with identities confirmed with a smaller precision. 

For all the galaxies we computed their physical properties, in particular their total infrared luminosities, $L_{TIR}$, making use of the CIGALE tool \citep{noll2009}.

In Fig.~\ref{fig:hist} we present redshift histograms $N(z)$ of our six galaxy samples selected at $\lambda$ from 2$\mu$m to 15$\mu$m. Similarly to the 24$\mu$m-selected sample analyzed by \citet{solarz2015}, in all the cases we can distinguish two up to three peaks: one with a maximum at $z\sim 0.5$, a second one at $z\sim1.2$, and the third one (sometimes tentative) at $z\sim2$. Because of the small statistics of the third peak we do not consider this highest $z$ sample here, but instead we divide the sample at $z < 1$, covering a very broad redshift range, into two samples corresponding to lower ($z\sim 0.4$) and higher ($z\sim 0.8$) redshifts.

CIGALE-based analysis of galaxy properties indicates that these samples are in all cases actually rather mixtures of galaxies of different properties. In the NIR filters (2-4$\mu$m) at the lower redshift we are dealing with a mixture of red galaxies whose emission may originate from 1.6 $\mu$m and 2.1 $\mu$m lines from cool stars, and star forming galaxies, while at higher $z$ these filters trace star forming galaxy population, with some LIRGs and ULIRGs. Emission from galaxies seen at all redshifts at 7 $\mu$m can be well explained as a mixture of red galaxies with old stellar populations and star forming galaxies. At yet longer wavelengths we start to see purer star forming populations, most likely traced by different PAH emission lines (7.3$\mu$m at lower $z$ and 3.3$\mu$m at higher $z$ for S11 and 6.2$\mu$m at lower $z$ and 12.7$\mu$m for L15), with some component of LIRGs and ULIRGs.

In Fig.~\ref{fig:lum} we see that the median $L_{TIR}$ of our galaxies changes both between samples and with redshift. As shown in the left panel, in all redshift bins NIR filters trace less luminous in the IR galaxy population than the MIR filters (which is consistent with the fact that NIR-selected samples contain some red passive galaxies which fade away at longer wavelengths and do not contribute significantly to $L_{TIR}$). At the lowest redshift this difference between the NIR and MIR filters is only weakly pronounced, but it becomes stronger at $z\sim 0.8$, and even more at $z \sim 1.2$. It can be also seen that at each redshift the three NIR filters follow populations of a very similar $L_{TIR}$ (of course, in a large percent they are the same objects), and the same observation can be made for a population traced by four MIR filters. As shown in the right panel,  $L_{TIR}$ of all our populations rises with redshift - which is an expected effect in a flux limited catalogue. However, the rise in median $L_{TIR}$ is stronger for the MIR-selected samples than for the NIR-selected ones, which may indicate that the longer wavelengths trace a more evolving population, which adds to the observational selection effect.

\begin{figure}[!ht]
     \plottwo{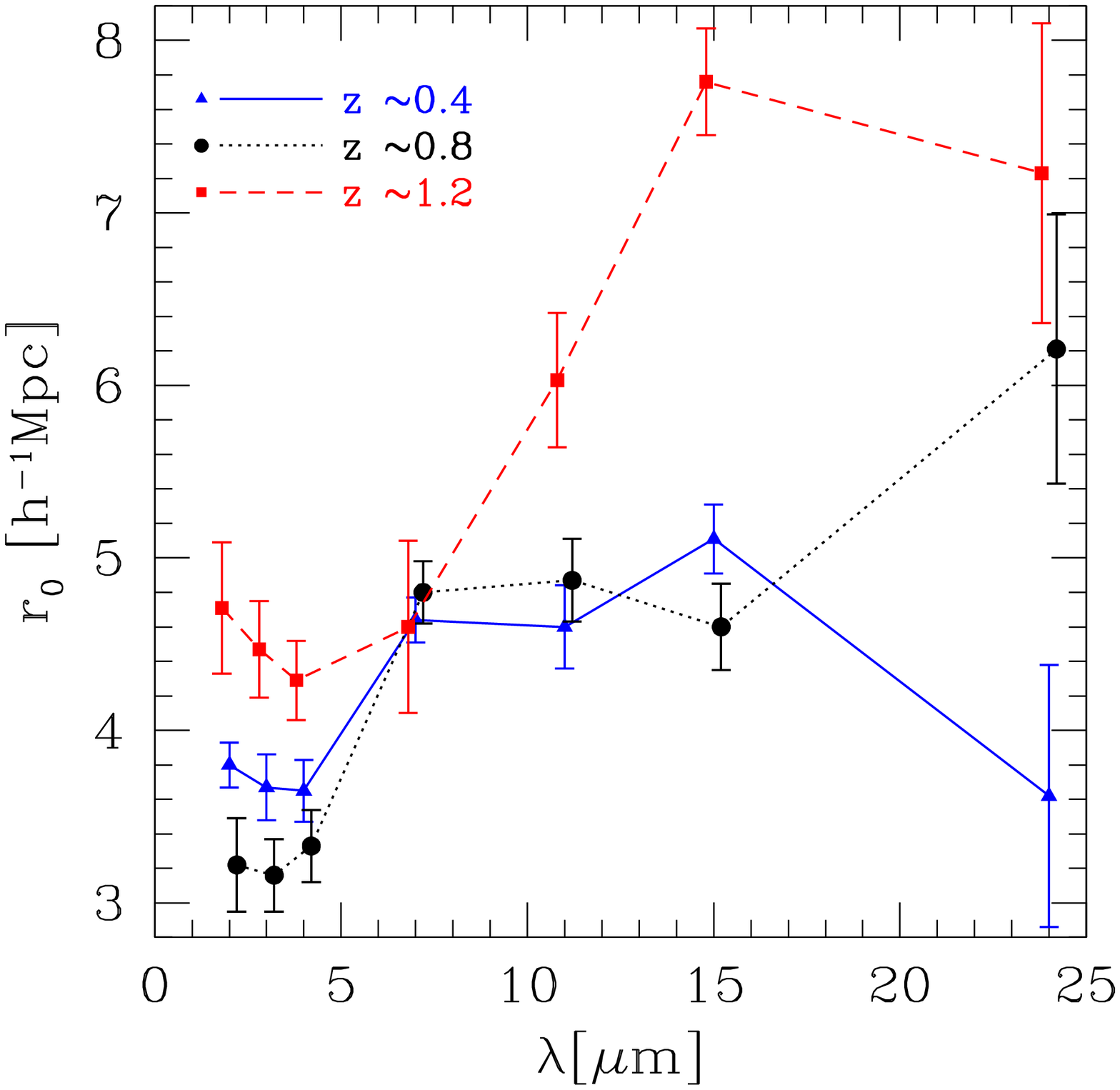}{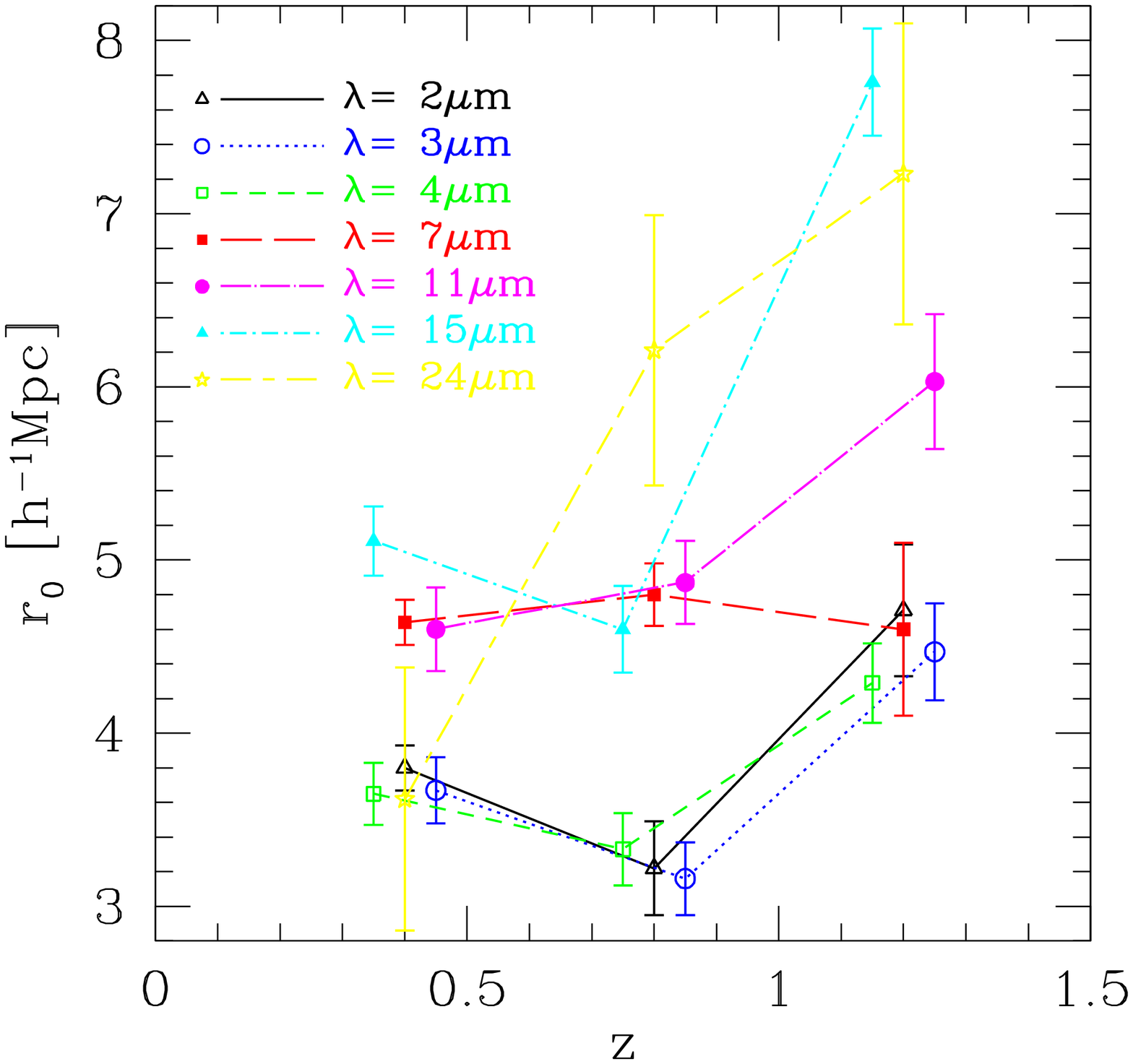}
\caption{Clustering length of samples of AKARI-NEP galaxies at different redshifts as a function effective wavelengths of the IRC filter (left panel) and as a function of redshift for different IRC filters (right panel).
}\label{fig:r0}
\end{figure}

\section{Galaxy clustering through AKARI IRC filters} 

The most commonly used statistical tool to study galaxy clustering is the two-point correlation function, real space $\xi(r)$ or angular $\omega(\theta)$. It measures an excess probability of finding a galaxy within a certain spatial distance $r$ or angular distance $\theta$ from another galaxy with respect to the random distribution. In practice, it can often be approximated by a power law $\xi(r) = (r/r0)^{\gamma}$. To compute the angular correlation function, we use the Landy \& Szalay estimator \citep{ls1993}, and then, knowing $N(z)$, we apply the Limber inversion to infer the real-space correlation function parameters. Here we present the (comoving) correlation length $r_{0}$ computed under assumption that the slope $\gamma$ is constant with a canonical value -1.8. 

The correlation length $r_{0}$ measured for different NEP-Deep samples is presented in Fig.~\ref{fig:r0}. We can see that at all redshifts MIR-selected galaxies are more strongly clustered than the NIR-selected ones. This implies that MIR-selected galaxies, albeit star forming, are located in relatively dense environments and massive dark matter haloes, while NIR-selected samples, despite of a component of red passive galaxies, are rather composed of field galaxies residing in less massive haloes. In spite of the obvious luminosity selection resultant from the catalogs being flux limited, discussed above, populations found at higher redshifts are not always more clustered. In particular, the NIR-selected population is less clustered at $z\sim 0.8$ than it is at lower and higher redshift - which indicates that we are dealing with a rather complex interplay between selection effects, galaxy evolution and genuine growth of large scale structure. MIR-selected galaxies are usually more clustered at higher $z$. There is an interesting case of 7$\mu$m selected population, which is almost equally clustered at all redshift ranges. It may partially result from the fact that it actually traces a similar galaxy mixture at different redshifts but most likely in this case all the effects mentioned above cancel out.

We conclude that the composition of a galaxy population strongly depends on the selection NIR or MIR filter, not only in terms of galaxy average properties but also in terms of clustering. MIR-selected galaxies are more strongly clustered than the NIR-selected ones at all probed redshifts. The MIR-bright dusty star forming galaxies, especially those located at higher redshifts, located in massive dark matter haloes (based on \citet{sheth1999} model we can estimate their masses to be $>5\times10^{12}$M$_{\odot}$) can belong to a fast evolving population which soon after the epoch at which we observe them reduced their star formation and ended as today's elliptical galaxies.


\subsection*{Acknowledgments}
This research is based on observations with AKARI, a JAXA project with the participation of ESA. It was supported by the Polish National Science Center grants  UMO-2012/07/D/ST9/02785 (AP), and UMO-2015/16/S/ST9/0043 (AS).



\end{document}